\documentclass[aps,prd,preprintnumbers,showpacs,twocolumn,groupedaddress,nofootinbib]{revtex4}
\usepackage{graphicx}
\usepackage{latexsym}
\def\beq{\begin{equation}}
\def\eeq{\end{equation}}
\def\bey{\begin{eqnarray}}
\def\eey{\end{eqnarray}}

\newcommand{\gev}{{\rm GeV}}

\newcommand{\mev}{{\rm MeV}}
\def\lsim{\mathrel{\raise.3ex\hbox{$<$\kern-.75em\lower1ex\hbox{$\sim$}}}}
\def\gsim{\mathrel{\raise.3ex\hbox{$>$\kern-.75em\lower1ex\hbox{$\sim$}}}}

\begin{document}

\title{Dark Forces and Light Dark Matter}  
\author{Dan Hooper$^{1,2}$, Neal Weiner$^3$, and  Wei Xue$^4$}
\affiliation{$^1$Center for Particle Astrophysics, Fermi National Accelerator Laboratory, Batavia, IL 60510, USA}
\affiliation{$^2$Department of Astronomy and Astrophysics, University of Chicago, Chicago, IL 60637, USA}
\affiliation{$^3$Center for Cosmology and Particle Physics, Department of Physics, New York University, New York, NY 10003, USA}
\affiliation{$^4$Department of Physics, McGill University, 3600 Rue University, Montreal, Quebec H3A 2T8, Canada}

\date{\today}

\begin{abstract}

We consider a simple class of models in which the dark matter, $X$, is coupled to a new gauge boson, $\phi$, with a relatively low mass ($m_{\phi} \sim 100 \, \mev -3 \, \gev$). Neither the dark matter nor the new gauge boson have tree-level couplings to the Standard Model. The dark matter in this model annihilates to $\phi$ pairs, and for a coupling of $g_X \approx 0.06 \times (m_X/10\,\gev)^{1/2}$ yields a thermal relic abundance consistent with the cosmological density of dark matter. The $\phi$'s produced in such annihilations decay through a small degree of kinetic mixing with the photon to combinations of Standard Model leptons and mesons. For dark matter with a mass of $\sim$10 GeV, the shape of the resulting gamma-ray spectrum provides a good fit to that observed from the Galactic Center, and can also provide the very hard electron spectrum required to account for the observed synchrotron emission from the Milky Way's radio filaments. For kinetic mixing near the level naively expected from loop-suppressed operators ($\epsilon \sim 10^{-4}$), the dark matter is predicted to scatter elastically with protons with a cross section consistent with that required to accommodate the signals reported by DAMA/LIBRA, CoGeNT and CRESST-II.

\end{abstract}

\maketitle

\section{Introduction}

Recent years have seen a surge of interest in light dark matter candidates. While such particles are not conventionally found in many of the most popular models (such as the Minimal Supersymmetric Standard Model, for example), a  number of reported observations have been interpreted as possible indirect and direct signals of dark matter particles with a mass of approximately 10 GeV~\cite{case}. These signals include the spectrum and angular distribution of gamma-rays from the Galactic Center as observed by the Fermi Gamma-Ray Space Telescope~\cite{gc},\footnote{Although astrophysical origins for the Galactic Center gamma-ray emission have also been proposed~\cite{Boyarsky:2010dr,pulsars}, the highly concentrated spatial morphology of this observed emission is difficult to accommodate in such scenarios~\cite{Linden:2012iv,gc}. In contrast, a dark matter distribution of $\rho(r)\propto r^{-1.3}$ provides a good fit to the observed gamma-ray signal and is in good agreement with expectations from state-of-the-art hydrodynamical simulations~\cite{ac}. 
} the synchrotron emission from the Milky Way's radio filaments~\cite{filaments}, and the diffuse synchrotron emission from the Inner Galaxy~\cite{timhaze} (known as the ``WMAP Haze''~\cite{wmaphaze}, whose presence has recently been confirmed by the Planck Collaboration~\cite{planck}). Observations reported from the direct detection experiments DAMA/LIBRA~\cite{dama}, CoGeNT~\cite{cogent}, and CRESST-II~\cite{cresst} 
 have also each been shown to possibly originate from the elastic scattering of approximately 10 GeV dark matter particles~\cite{arXiv:1110.5338,Fox:2011px,Kopp:2011yr}. And while a number of null results have been presented as a challenge to this direct detection evidence~\cite{Aprile:2011hi,Angle:2011th,Ahmed:2010wy,Ahmed:2012vq} (see also, however, Refs.~\cite{juancdms,Collar:2011kf,Collar:2010ht,Collar:2010gg}), it has become clear that the breadth of possibilities for light WIMPs is much larger than previously appreciated.

If dark matter particles are in fact responsible for this collection (or any subset of this collection) of direct and indirect signals, they must possess a number of rather specific properties. In particular, in order to explain all of these observations with a single species of dark matter particles (with a mass of approximately 10 GeV), the following requirements must be satisfied:  
\begin{itemize}
\item{To accommodate the shape of the gamma-ray spectrum observed from the Galactic Center~\cite{gc}, dark matter annihilations must not proceed primarily to quarks, but to final states such as $\tau^+\tau^-$. The gamma-ray spectrum tentatively reported from the Virgo Cluster also possesses similar features~\cite{virgo}. We will show later in this article that dark matter annihilations to mesons (including neutral pions) or to $e^+ e^- \gamma$ can also provide a good fit to the measured spectrum.}
\item{To produce the distinctive spectrum of synchrotron emission that is observed from the Milky Way's non-thermal radio filaments~\cite{filaments}, dark matter annihilations must inject an extremely hard spectrum of electrons (sometimes described in the radio literature as ``monoenergetic''~\cite{1988A&A...200L...9L,1992A&A...264..493L}). Dark matter which annihilates to $e^+ e^-$ a significant fraction of the time can accommodate both the observed characteristics of the radio filaments, as well as the WMAP/Planck Haze~\cite{timhaze}, and could also potentially account for much of the excess isotropic radio background~\cite{radiobg}.}
\item{The total cross section required to normalize the annihilation rate to the observed gamma-ray and radio fluxes is approximately $\sim$$10^{-26}$ cm$^3$/s, although uncertainties in the dark matter distribution make the extraction of this quantity uncertain at the level of a factor of a few. This value is strikingly similar to that required to thermally produce the measured abundance of dark matter in the early universe ($\sigma v \simeq 3 \times 10^{-26}$ cm$^3$/s).}
\item{The spectra and time variation of events reported by the DAMA/LIBRA, CoGeNT and CRESST-II collaborations collectively favor a spin-independent elastic scattering cross section between dark matter and nucleons on the order of $\sigma \sim 10^{-41}$ cm$^2$ (assuming equal couplings to protons and neutrons)~\cite{arXiv:1110.5338,Fox:2011px,Kopp:2011yr}.}
\end{itemize}

And while some of the features listed above are not found among many of the most popular dark matter candidates (such as neutralinos), various models satisfying these requirements have been proposed~\cite{Buckley:2010ve}. Perhaps the simplest scenario considered thus far is one in which the dark matter annihilates to the desired charged lepton final states through the exchange of a new gauge boson with much larger couplings to leptons than to quarks. Such a leptophilic gauge boson could arise from the addition of a new gauge group, such as the anomaly free $U(1)_{L_i-L_j}$, for example. Any gauge boson that couples to electrons (as required to generate the synchrotron spectrum observed from radio filaments), however, must contend with the rather stringent constraints from LEP II. In particular, in order for the dark matter to annihilate through the exchange of a leptophilic gauge boson at a rate high enough to avoid being overproduced in the early universe while also avoiding the constraints from LEP II requires either that the gauge boson couples much more strongly to the dark matter than to electrons, or that the mass of the gauge boson lies near the resonance $m_{Z'} \sim 2 m_{X}$, where $m_X$ denotes the mass of the dark matter candidate~\cite{Buckley:2010ve}. And while either of these possibilities represent viable options from a model building standpoint, neither are what one might have naively expected nature to provide.

In this article, we consider an alternative class of dark matter models capable of explaining the indirect and direct signals described above. Again, we consider a new gauge boson, but with a mass lighter than that of the dark matter itself, $m_{\phi} < m_X$. If the gauge group responsible for this new gauge boson is charged only to the dark matter, such as $U(1)_X$, then dark matter annihilations will produce pairs of the new boson, which then decay through kinetic mixing with the photon to Standard Model states. As we will show, for very plausible values of the gauge coupling ($g_X\approx 0.06$), gauge boson mass ($m_{\phi} \sim$ 100 MeV- 3 GeV), and degree of kinetic mixing ($\epsilon \sim 10^{-3}-10^{-6}$), the dark matter in this model can account for the observed gamma-ray and synchrotron spectra, as well as the anomalous direct direction signals.


\section{Dark Matter Annihilation Through A New Dark Force}
\label{anni}
Dark matter interacting through dark forces has been widely discussed in recent years, especially within the context of efforts to provide an explanation for the PAMELA positron excess \cite{ArkaniHamed:2008qn,Pospelov:2008jd,fsr,Cholis:2008qq,Kang:2010mh}. The idea that dark matter might be charged under a $U(1)$ that kinetically mixes with the photon was first considered by Holdom nearly three decades ago~\cite{Holdom:1985ag}. Models in which the dark matter could freeze-out by annihilations into a light metastable dark force carrier were considered much more recently by the authors of Ref.~\cite{Finkbeiner:2007kk}, who noted that high energy $e^+e^-$ final states were a natural consequence of this channel. Such signals were subsequently studied in Ref.~\cite{Cholis:2008vb}. Such models were studied in general in Ref.~\cite{Pospelov:2007mp}, which examined both heavy WIMPs as well as the possibility of $\sim$MeV mass WIMPs to explain the 511 keV line observed by INTEGRAL.

Within the context of light WIMPs, Refs.~\cite{Morrissey:2009ur,Chang:2010yk,Cohen:2010kn} pointed out that a $\sim$\,GeV mass $U(1)$ gauge boson which kinetically mixes with electromagnetism could lead to a large elastic scattering cross section between dark matter and nuclei. And while the leptonic phenomenology (i.e., PAMELA) has been well explored for dark forces in the case of heavy WIMPs, the indirect signals for the slightly heavier $\phi$ (with associated hadronic cascades) have not been as thoroughly studied. Moreover, within the context of light WIMPs, and specifically with connections to observations of the Galactic Center, the associated gamma-ray phenomenology has not previously been explored.

This simple model we consider in this article consists of a stable Dirac fermion, $X$, which will serve as our dark matter candidate, and a new $U(1)_X$ gauge group, broken to provide a massive vector boson, $\phi$. If the mass of the gauge boson is much lighter than the mass of the dark matter candidate, $m_{\phi} \ll m_X$, then dark matter annihilations will proceed dominantly through the $t$-channel exchange of an $X$ to a pair of $\phi$ particles with a cross section given by~\cite{Cholis:2008vb}:
\begin{equation}
\sigma v_{XX \rightarrow \phi \phi} \simeq \frac{\pi \alpha^2_X}{m^2_{X}} \approx 3\times 10^{-26} \, {\rm cm}^3/{\rm s} \, \bigg(\frac{g_X}{0.06}\bigg)^4  \bigg(\frac{10\,{\rm GeV}}{m_X}\bigg)^2,
\end{equation}
where $\alpha_X \equiv g^2_X/4\pi$ and $g_X$ is the gauge coupling of the dark force. Note that for dark matter particles with a mass in the range motivated by the aforementioned indirect and direct signals ($m_X\sim 10$ GeV), the measured cosmological density of dark matter will be produced thermally in the early universe for a gauge coupling of $g_X \approx 0.06$, regardless of the mass of the light force carrier, $m_{\phi}$. With this in mind, we will fix the gauge coupling to this value throughout the remainder of this paper.

The leading interaction between the Standard Model and the dark sector is kinetic mixing between the photon and the $\phi$, $\mathcal{L}=\frac{1}{2}\epsilon F'_{\mu \nu} F^{\mu \nu}$.\footnote{Mixing between the $\phi$ and the Standard Model $Z$ is also possible, but is expected to be suppressed by $\sim m^2_{\phi}/m^2_{Z}$ relative to that with the photon.} This has the effect of inducing effective couplings between the $\phi$ and the particle content of the Standard Model, proportional to their electric charge. There is no robust prediction for the size of this coupling -- in the effective theory, any value of $\epsilon$ is technically natural (see the discussion in e.g., Ref.~\cite{ArkaniHamed:2008qp}). If the Standard Model is embedded in a Grand Unified Theory (GUT), however, this coupling can only be generated after GUT breaking at the loop level. Such a loop of heavy states carrying both hypercharge and $X$ gauge charge naturally leads to kinetic mixing at the following order~\cite{Holdom:1985ag,Baumgart:2009tn,Cohen:2010kn}:
\begin{equation}
\epsilon \sim \frac{g_X g_Y \cos \theta_W}{16 \pi^2} \, \log\bigg(\frac{M'}{M}\bigg) \sim 1.2 \times 10^{-4} \, \log\bigg(\frac{M'}{M}\bigg),
\label{epsilon}
\end{equation}
where $(M'/M)$ is the ratio of the masses in the loop. Thus we expect the kinetic mixing to occur at the level of $\epsilon\sim 10^{-4}$ or less, modulo the possibility of a large hierarchy between $M'$ and $M$. Furthermore, if the splitting between the different components of the GUT multiplet is generated at loop order, then $\epsilon$ becomes further suppressed by two loops. A similar set of arguments can be applied if the dark $U(1)$ is embedded into its own non-Abelian group, at which point 3- and 4-loop suppression becomes natural. Consequently, very small values for $\epsilon$ could be possible. We can place a lower limit on $\epsilon$, however, by requring that it be large enough to thermalize the system through the process $f \gamma \leftrightarrow f \phi$. This requires that $T^2/M_{\rm Pl} = \alpha^2 \epsilon^2 T$ (for $T \gg m_\phi, m_f$). Thus, for $\epsilon \gsim 10^{-7}$ the system should be thermalized before the temperature of WIMP decoupling.

After a dark matter annihilation produces a pair of $\phi$ particles, those particles will decay via this small kinetic mixing into Standard Model states.  The dominant decay channels of the $\phi$ depend on its mass. For $2 m_e < m_{\phi} < 2 m_{\mu}$, $\phi$ decays proceed almost entirely to $e^+ e^-$, whereas for $2 m_{\mu} < m_{\phi} \lsim$ a few hundred MeV, $\phi$ decays produce a combination of $e^+ e^-$ and $\mu^+ \mu^-$. For $\phi$'s with masses between a few hundred MeV and a few GeV, decays proceed to a combination of charged leptons and mesons. Above a few GeV, $m_{\phi} \gg \lambda_{\rm QCD}$, and the $\phi$ decays directly to quark-antiquark pairs (along with charged lepton pairs).



\begin{figure}[t]
\includegraphics[angle=0.0,width=3.3in]{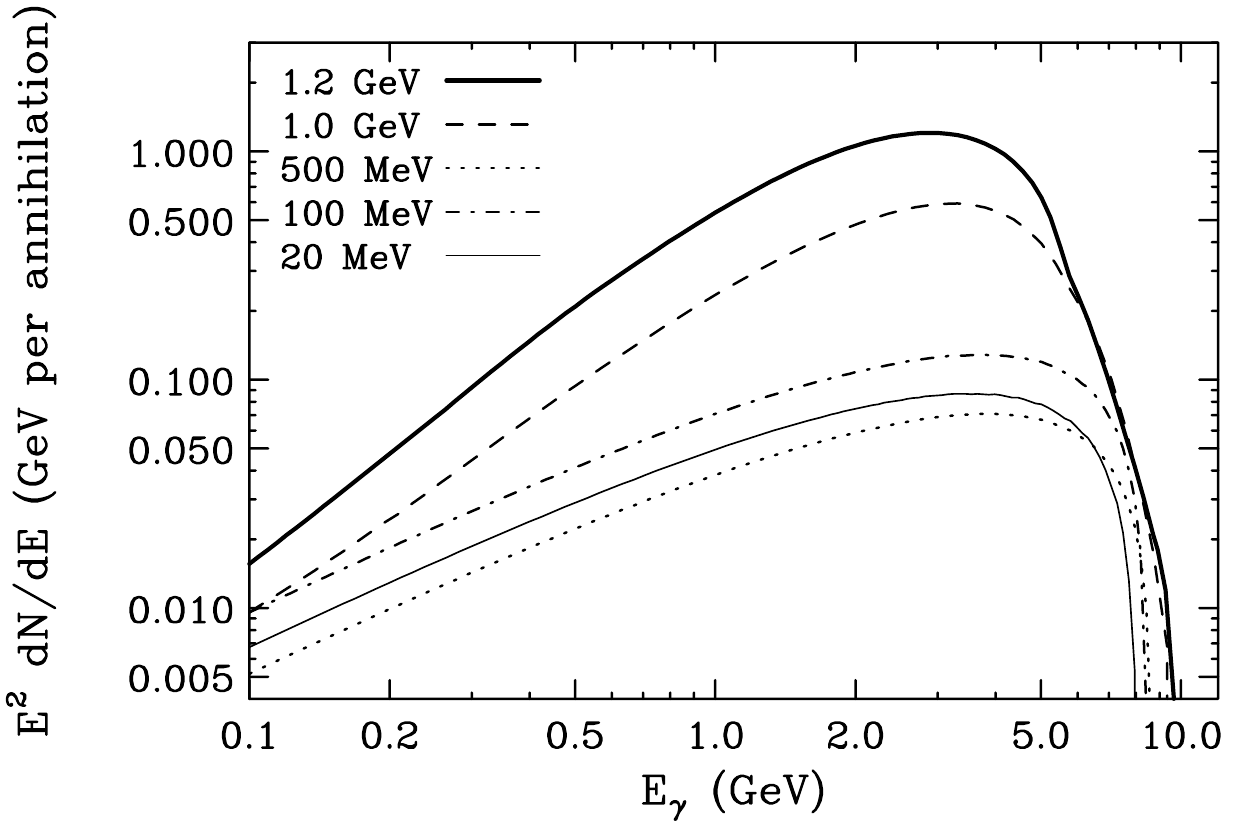}
\includegraphics[angle=0.0,width=3.3in]{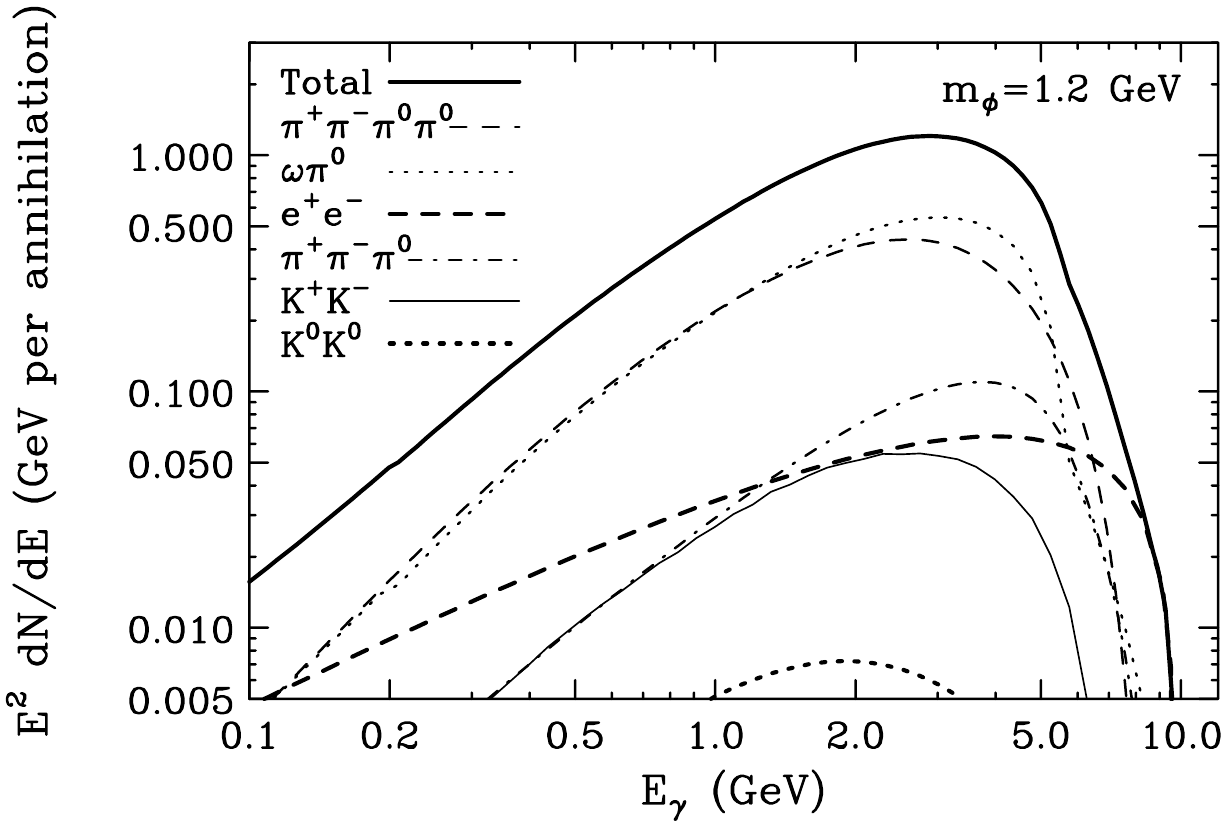}
\caption{Upper: The gamma-ray spectrum from dark matter annihilations to two dark gauge bosons, for various choices of the dark gauge boson's mass. Lower: Leading contributions to the gamma-ray spectrum for the case of $m_{\phi}=1.2$ GeV. For $m_{\phi} \lsim 700$ MeV, the gamma-ray spectrum is dominated by final state radiation from $\phi$ decays to $e^+ e^-$. For heavier values of $m_{\phi}$, decays to mesons provide the most significant contributions.}
\label{spec}
\end{figure}

\begin{figure}
\includegraphics[angle=0.0,width=3.3in]{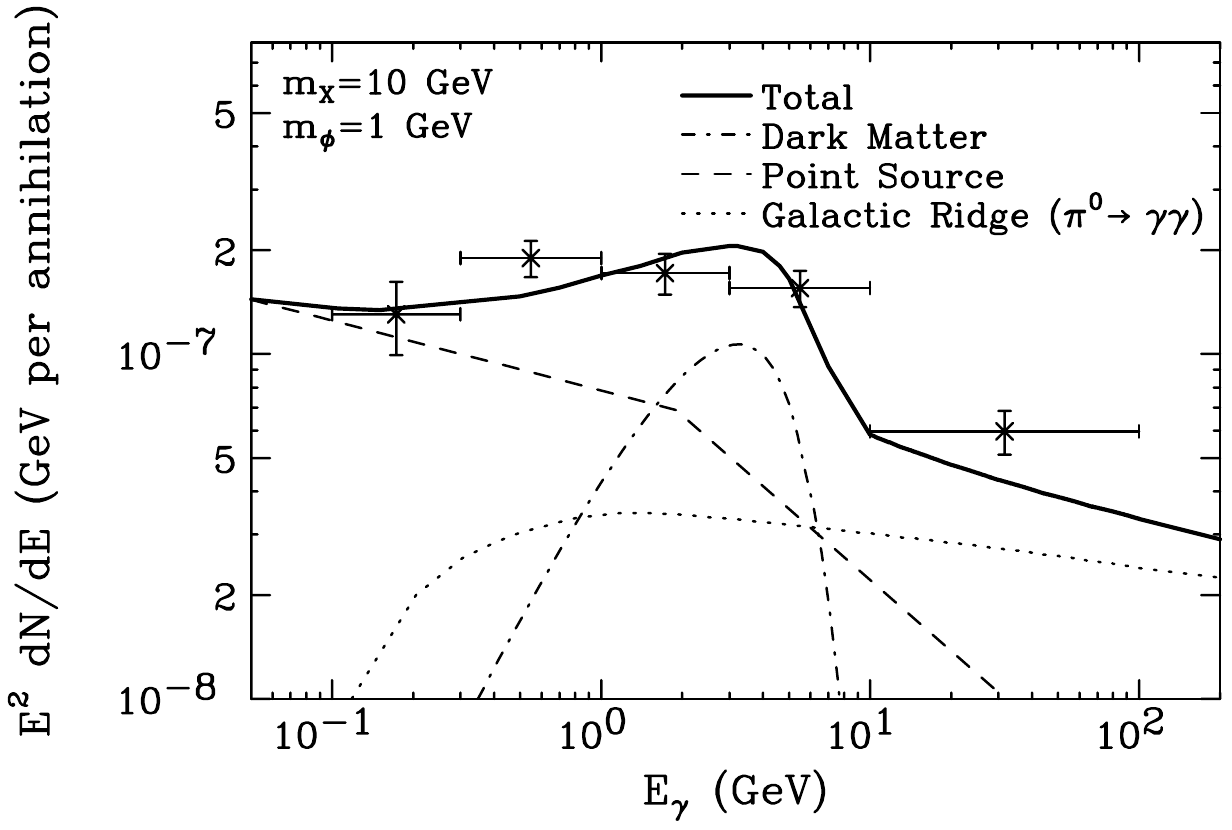}
\includegraphics[angle=0.0,width=3.3in]{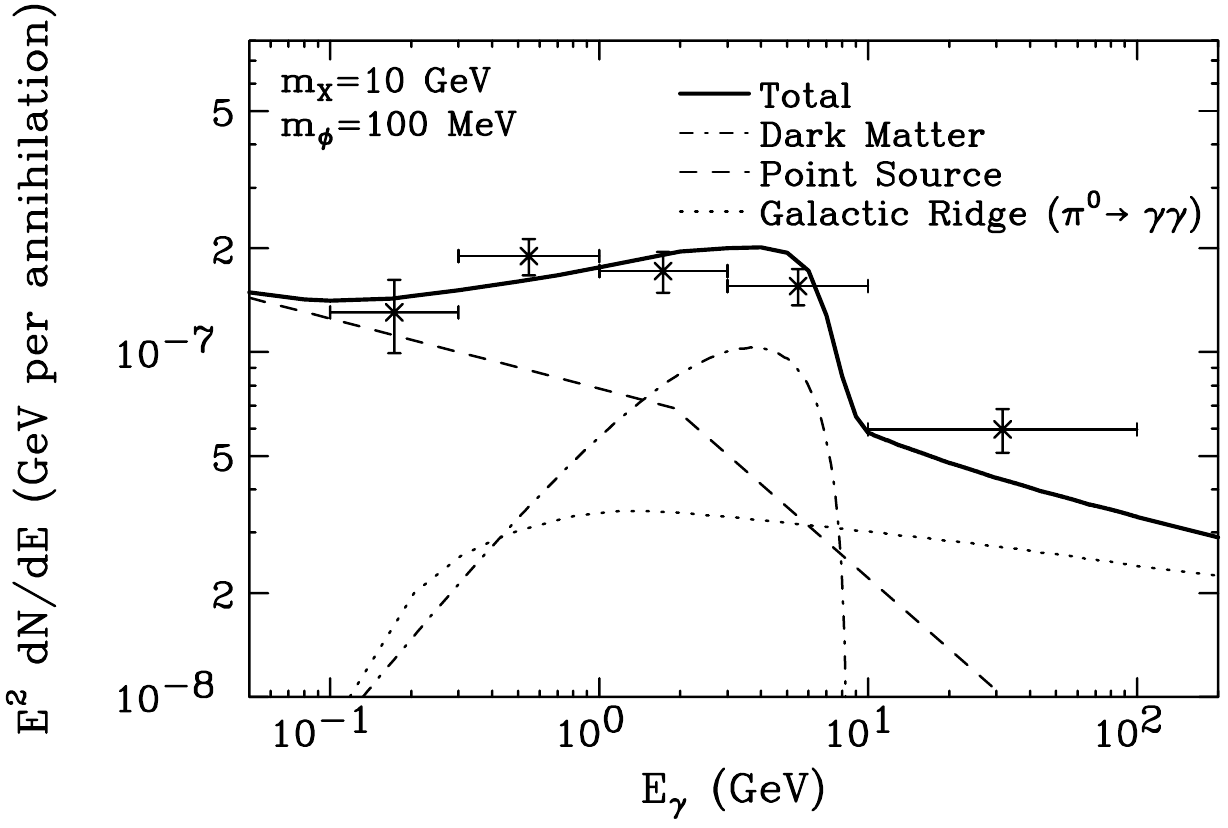}
\caption{The spectrum of gamma-rays from the Galactic Center observed by the Fermi Gamma-Ray Space Telescope (error bars)~\cite{gc} compared to that predicted from dark matter annihilations in the model presented here (dot-dashed), along with the measured emission from the central point source (dashed) and emission from the Galactic Ridge as extrapolated from higher energy HESS data (dots). In the upper and lower frames, we have considered dark gauge bosons with masses of 1 GeV and 100 MeV, respectively. In each case, we have fixed the gauge coupling to provide a total annihilation cross section of $\sigma v=3\times 10^{-26}$ cm$^3$/s, as required to produce the measured abundance of dark matter in the early universe. We have adopted a halo profile consistent with the observed morphology of the extended gamma-ray emission, $\rho(r)\propto r^{-1.3}$, normalized such that the dark matter density in the local neighborhood is 0.17 GeV/cm$^3$ and 0.35 GeV/cm$^3$ in the upper and lower frames, respectively.}
\label{fermi}
\end{figure}

The shape of the gamma-ray spectrum produced in dark matter annihilations therefore depends on the mass of the $\phi$. In the top frame of Fig.~\ref{spec}, we show the gamma-ray spectrum per dark matter annihilation for several values of $m_{\phi}$. For values below $\sim$1 GeV, the gamma-ray spectrum is dominated by final state radiation associated with the process $XX\rightarrow \phi \phi$, $\phi\rightarrow e^+ e^-$, boosted as described in Ref.~\cite{fsr}. For heavier masses ($m_{\phi} \sim$1-3 GeV), decay channels such as $\pi^+\pi^-\pi^0\pi^0$, $\omega \pi^0$, $K^+ K^-$, $\pi^+ \pi^- \pi^0$, and $K^0 K^0$ each contribute significantly to the resulting gamma-ray spectrum.\footnote{The branching fractions of a particle decaying through kinetic mixing with the photon can be determined using the measurements compiled at http://durpdg.dur.ac.uk/hepdata/online/rsig/index.html. The photon spectrum has been calculated based the effective Lagrangian approach and chiral perturbation theory~\cite{photoncalc}.} The leading contributions to the gamma-ray spectrum are shown in the bottom frame of Fig.~\ref{spec} for the case of $m_{\phi}=1.2$ GeV (with branching fractions of the $\phi$ as given in Ref.~\cite{Meade:2009rb}). In the $m_{\phi}=$1 GeV case, decays to $\pi^+\pi^-\pi^0$, $\omega \pi^0$, $\pi^+\pi^-\pi^0\pi^0$ and final state radiation associated with decays to $e^+ e^-$ each contribute significantly to the gamma-ray spectrum, with branching fractions of approximately 20\%, 2\%,  1.5\% and 33\%, respectively.

In Fig.~\ref{fermi}, we compare the spectrum of gamma-rays predicted in this model to the spectrum from the Galactic Center, as observed using the Fermi Gamma-Ray Space Telescope~\cite{gc}, for two choices of $m_{\phi}$. We have normalized the gamma-ray flux using a gauge coupling which yields an annihilation cross section of $3\times 10^{-26}$ cm$^3$/s and have adopted a dark matter distribution consistent with the observed morphology of the gamma-ray signal ($\rho\propto r^{-1.3}$, normalized such that the dark matter density in the local neighborhood is 0.17 GeV/cm$^3$ and 0.35 GeV/cm$^3$ in the upper and lower frames, respectively). Along with the contribution from dark matter annihilations (dot-dashed), we include the measured spectrum of the central point source (dashes)~\cite{Boyarsky:2010dr,aharonian} and the emission extrapolated from higher energy observations of the Galactic Ridge (dotted)~\cite{ridge}. For each of these cases ($m_{\phi}=$100 MeV or 1 GeV), we find that dark matter annihilations can not only accommodate the spectral shape of the observed signal, and also automatically provide the approximate annihilation rate required to normalize the overall flux (once the dark matter distribution is fixed to the observed morphology of the gamma-ray signal).

The peculiar spectrum of synchrotron emission observed from the Milky Way's non-thermal radio filaments~\cite{filaments} can also be easily accounted for in this class of dark matter models. In particular, the significant branching fraction for $\phi \rightarrow e^+ e^-$ in this model leads to a spectrum of electrons and positrons that is sufficiently hard and which cuts off above $m_X$ sufficiently abruptly to account for the filaments' observed spectral characteristics. We also note that when compared to the case of dark matter particles which annihilate directly to charged leptons (democratically to each flavor), as considered in Ref.~\cite{filaments}, the class of models being considered here deposits a larger fraction of the total annihilation power into electrons and positrons. Quantitatively, for a value of $m_{\phi}=1$ GeV (100 MeV), the total power injected into electrons and thus into synchrotron emission is larger than in the democratic lepton benchmark model by a factor of 1.7 (2.8). This provides a somewhat better match to the required normalization of the observed filaments (see the discussion of filament widths in Sec.~4 of Ref.~\cite{filaments}). Similarly, this increased power into electrons makes it possible to account for the WMAP Haze with a somewhat lower magnetic field strength in the inner kiloparsecs of the Milky Galaxy ($\sim$$10$-$15$ $\mu$G at $\sim$1 kpc from the Galactic Center instead of $\sim$$20\mu$G that is otherwise required~\cite{timhaze,case}).

\section{Elastic Scattering}
\label{direct}

The kinetic mixing between the $\phi$ and the photon leads to spin-independent elastic scattering between the dark matter and protons, with a cross section that is given by:
\begin{eqnarray}
\sigma_{Xp} &=& \frac{g^2_2 \sin^2 \theta_{W} g^2_X \epsilon^2 m^2_X m^2_p}{\pi m^4_{\phi} (m_X+m_p)^2} \\
&\approx& 1.6\times 10^{-40} \, {\rm cm}^2 \, \bigg(\frac{\epsilon}{7\times 10^{-5}}\bigg)^2 \bigg(\frac{1\,{\rm GeV}}{m_{\phi}}\bigg)^4. \nonumber
\end{eqnarray}
As this cross section is generated through the photon's coupling to electric charge, the corresponding cross section with neutrons is negligible. With kinetic mixing of $\epsilon \sim 7 \times 10^{-5}$ ($7 \times 10^{-7}$), a 1 GeV (100 MeV) gauge boson will generate an elastic scattering cross section compatible with that required by the signals observed by DAMA/LIBRA, CoGeNT, and CRESST-II.\footnote{Since the commonly quoted number is cross section per nucleon (assuming equal couplings to protons and neutrons), one must scale this up by $A^2/Z^2$ to find the comparably required cross section per proton. For example, the nucleon-level cross section $\sigma_{NX} \sim (0.7-3)\times 10^{41}$ cm$^2$ required to accommodate the spectrum of events observed by CoGeNT~\cite{arXiv:1110.5338} translates to a cross section of $\sigma_{pX}\sim (0.4-1.6)\times 10^{-40}$ cm$^2$ with protons. Similarly, $\sigma_{NX} \sim (0.4-1.0)\times 10^{41}$ cm$^2$ required to accommodate CRESST translates to $\sigma_{pX}\sim (1.6-4.0)\times 10^{-40}$ cm$^2$ (for scattering with either oxygen or calcium targets.} When this is compared to the values of $\epsilon$ that are expected according to Eq.~\ref{epsilon}, we find that the anomalous signals reported by these three direct detection experiments are in good agreement with the scattering rates anticipated in the model under consideration, in particular for the case of $m_{\phi} \sim 1$ GeV.


\section{Constraints}

\begin{figure}
\includegraphics[angle=0.0,width=3.3in]{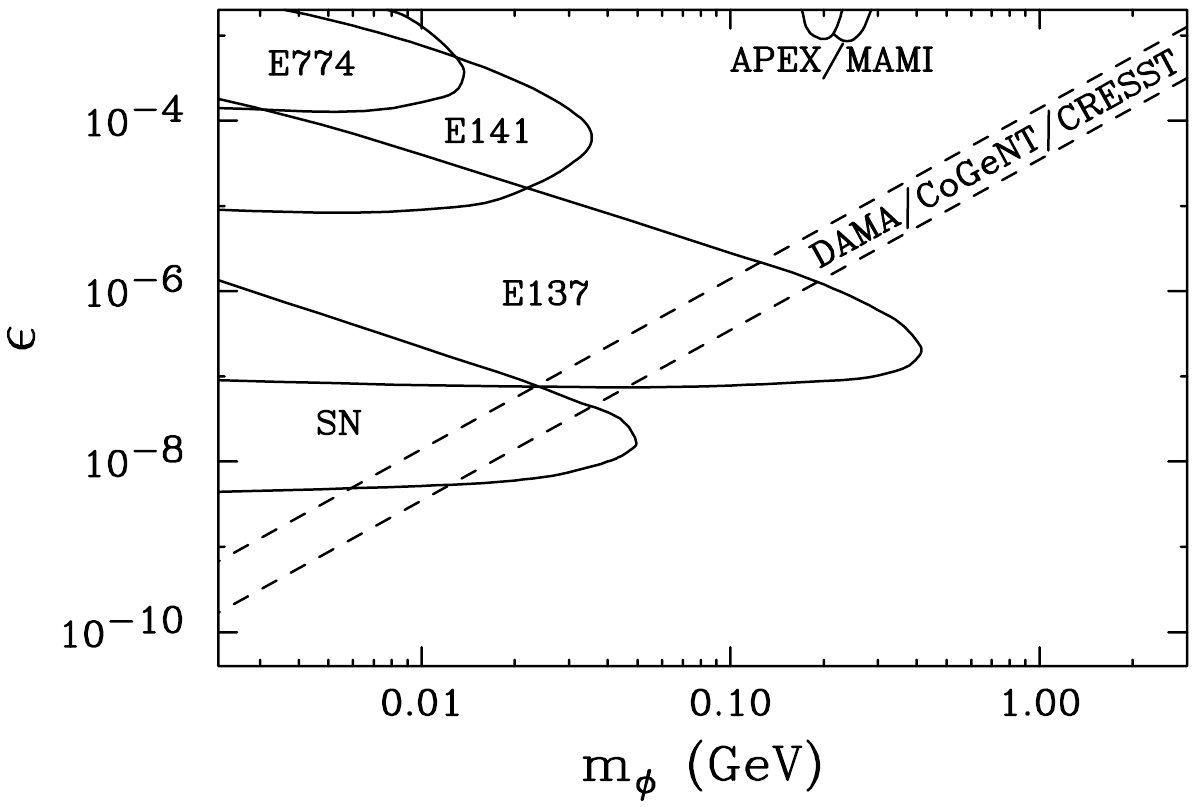}
\caption{Current laboratory and supernova (SN) constraints on the mass and kinetic mixing of the dark force carrier. Also shown is the approximate range of parameters required to accommodate the direct detection signals reported by the DAMA/LIBRA, CoGeNT and CRESST-II collaborations (see Sec.~\ref{direct}). For a summary of these constraints and prospects for future laboratory searches, see Ref.~\cite{intensityfrontier}.}
\label{intensity}
\end{figure}

Compared to other light dark matter candidates, dark matter in this class of models is relatively unconstrained. The most significant constraints on this scenario are related to the dark photon, $\phi$, rather than the dark matter itself. A broad range of searches for dark photons is described in Ref.~\cite{intensityfrontier} (see also Ref.~\cite{Gninenko:2012eq}). In Fig.~\ref{intensity} we have summarized the current status of laboratory constraints on such particles, as well as those derived from supernovae. Also shown is the approximate range of parameters required to account for the elastic scattering cross section implied by DAMA/LIBRA, CoGeNT, and CRESST-II (see Sec.~\ref{direct}). 

From this figure, we see that the parameter space favored by these direct detection anomalies is unconstrained for masses above $m_{\phi} \sim 100 \, \mev$. We also note that the values of $\epsilon$ favored by the one-loop calculation of Eq.~\ref{epsilon} ($\epsilon \sim 10^{-4}$) fall safely within this allowed region. If the degree of kinetic mixing between the $\phi$ and the photon is further suppressed, another seemingly viable window appears at $m_{\phi} \sim 1-10 \, \mev$. Constraints from the ellipticity of dark matter halos, however, require that $m_{\phi} \gsim 30$ MeV for the case being considered here~\cite{haloshapes}. Furthermore, as stated in Sec.~\ref{anni}, values of $\epsilon$ less than $\sim10^{-7}$ are insufficient to thermalize the dark matter prior to decoupling in the early universe.\footnote{Note that for $m_{\phi} \lsim 10$ MeV (below the typical energy that is exchanged in scattering between dark matter and nuclei), the direct detection cross section becomes saturated. The details of this depend on the energy threshold and target material utilized by the experiment.~\cite{saturate}} For these reasons, we focus on the $m_{\phi} \gsim 100$ MeV region of parameter space. Also note that although Sommerfeld enhancements would be expected to boost the annihilation rate in the Galactic halo at the level of roughly a factor of two in the $m_{\phi} \sim 1-10 \, \mev$ case, such effects are negligible for larger values of $m_{\phi}$~\cite{ArkaniHamed:2008qn}.


As interactions between the dark matter and the Standard Model are mediated by the light $\phi$ in this scenario, collider searches for mono-photons~\cite{Fox:2011fx}  or mono-jets~\cite{Beltran:2010ww,Bai:2010hh} plus missing energy are essentially doomed to failure. And although limits on GeV-scale neutrinos from dark matter annihilations in the Sun can significantly constrain other light WIMP models~\cite{Kappl:2011kz}, annihilations in this model produce neutrinos only through mesons which are stopped in the solar medium before they decay. Similarly, due to the light mass of the $\phi$, no antiproton cosmic rays are expected to be produced \cite{Finkbeiner:2007kk,Cholis:2008vb}. Gamma-ray constraints, such as those derived from observations of dwarf spheroidal galaxies~\cite{dwarfs}, still apply but are currently a factor of a few too weak to constrain this scenario~\cite{case}. Similarly, constraints from the cosmic microwave background~\cite{cmb} apply to this scenario, but again are not currently sensitive to this class of models (although Planck may be). 

There is some hope for discovery in future laboratory experiments, however. When dark force models are embedded into supersymmetric theories, there is a natural expectation of ``lepton jets'' \cite{ArkaniHamed:2008qp} at the Large Hadron Collider at the ends of sparticle cascades. Low energy searches such as APEX \cite{apex} and MAMI \cite{mami} have already begun to probe interesting regions of parameter space, while future experiments such as Darklight \cite{darklight} and HPS \cite{hps} are expected to probe broad ranges of open parameter space.

\section{Heavy Dark Matter and Photon Signals}

\begin{figure}
\includegraphics[angle=0.0,width=3.4in]{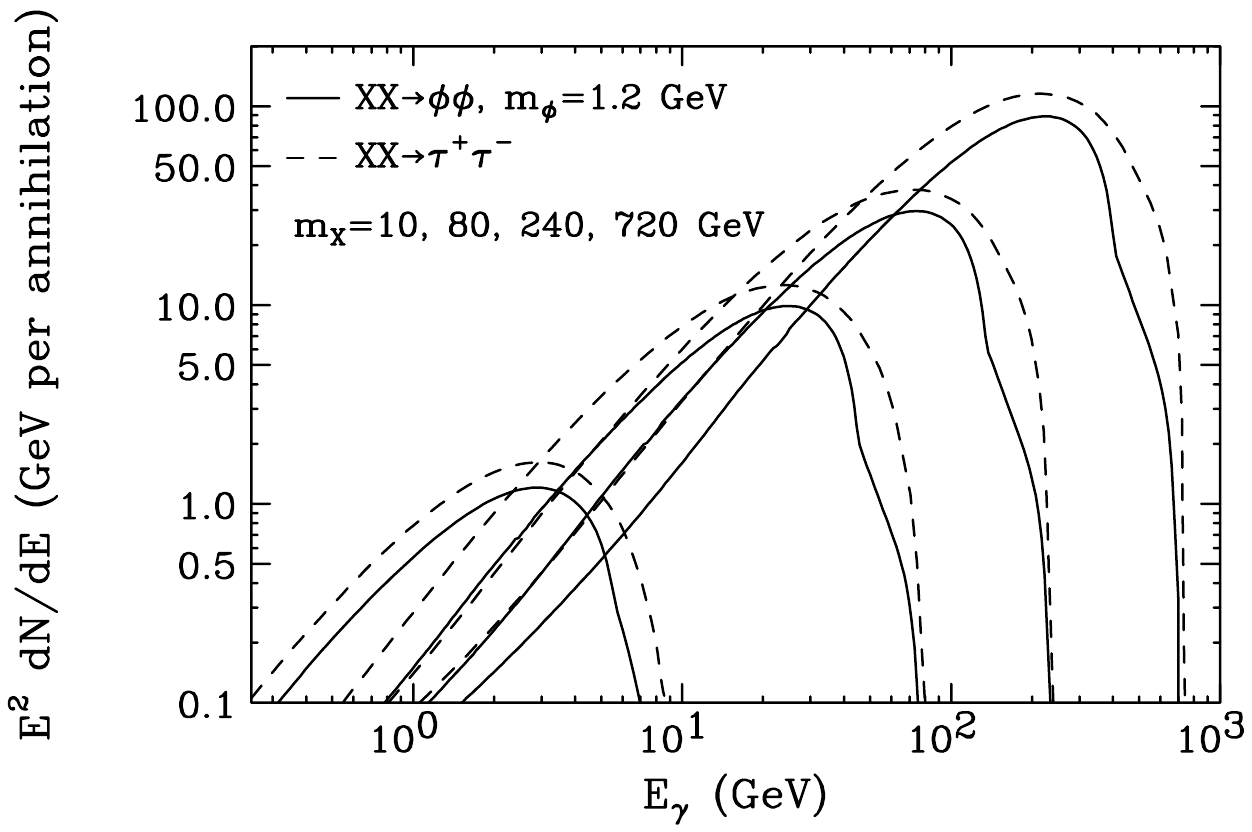}
\includegraphics[angle=0.0,width=3.4in]{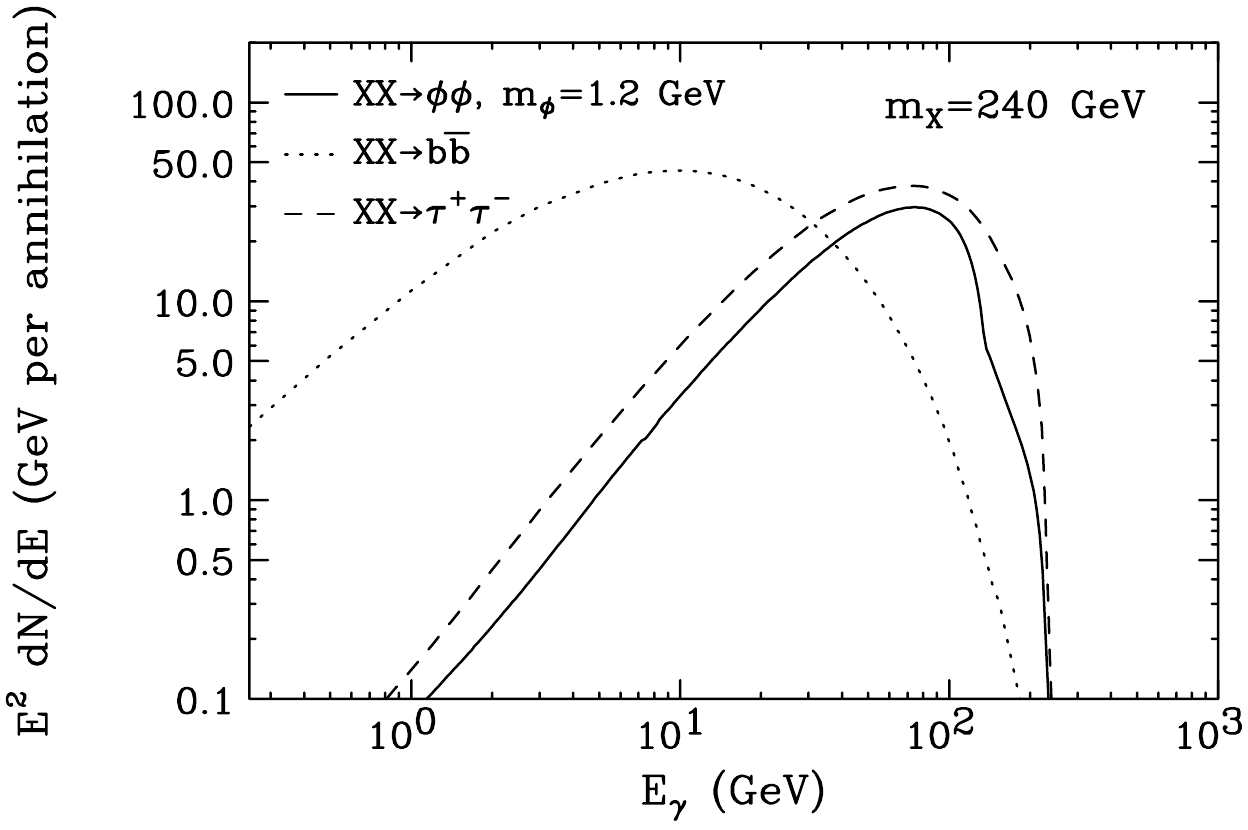}
\caption{The spectrum of gamma-rays from dark matter annihilations to a pair of 1.2 GeV dark force carriers compared to the spectrum from dark matter annihilations to $\tau^+ \tau^-$, for four choices of the dark matter mass (top), and compared to the spectrum from dark matter annihilations to $b\bar{b}$ or $\tau^+ \tau^-$ for $m_X=240$ GeV (bottom). The gamma-ray spectrum from $\phi$ decays is dominated by decays of mesons (especially decays to $\pi^+ \pi^- \pi^0 \pi^0$ and $\omega \pi^0$) and is similar to that resulting from $\tau^{\pm}$ decays.}
\label{fig:heavysignal}
\end{figure}

While we have focused up to this point on the signals of light WIMPs, it is also worth considering the consequences for heavier WIMPs interacting through a dark force. The leptonic signals (e.g., PAMELA) have been widely discussed previously, but the case with $m_\phi \gsim 1 \,\gev$ (for which $\phi$'s decay a significant fraction of the time to hadronic final states) has been less explored (see, however, Ref.~\cite{Meade:2009rb}). We show in the top frame of Fig.~\ref{fig:heavysignal} the gamma-ray spectrum from WIMPs of various masses, annihilating to $\phi$'s with a mass of 1.2 GeV. We directly compare this to the spectrum from dark matter of the same mass annihilating into $\tau^+ \tau^-$. We see that very similar spectra result from these two cases, although the gamma-ray flux in the dark forces case is suppressed by about 25\% relative to that predicted from annihilations to $\tau^+ \tau^-$. This similarity is not surprising when one considers that the gamma-ray spectrum from $\phi$ decays is dominated by decays to mesons, such as $\phi \rightarrow \pi^+ \pi^- \pi^0 \pi^0$ and $\phi \rightarrow \omega \pi^0$, whereas the gamma-ray spectrum from tau decays is dominated by the channels $\tau^{-} \rightarrow \nu_{\tau} \pi^- \pi^0$ and $\tau^{-} \rightarrow \nu_{\tau} \pi^- \pi^0 \pi^0$. In both of these cases, the highly boosted $\pi^0$'s lead to a very hard gamma-ray spectrum, especially when compared to that resulting from dark matter annihilations to quarks or gauge bosons (as can be seen in the bottom frame of Fig.~\ref{fig:heavysignal}.

\bigskip

\section{Summary and Conclusions}

Models in which the dark matter interacts through dark forces (i.e., forces without tree level couplings to the Standard Model) possess a number of interesting and distinctive phenomenological characteristics. Whereas previous studies of such models have focused on heavy WIMPs, in this article we have discussed the implications for direct and indirect detection of light WIMPs ($m_X\sim$10 GeV) which interact through a light dark force carrier ($m_{\phi}\sim$100 MeV-3 GeV).

Dark matter particles in this scenario annihilate to pairs of dark gauge bosons, $\phi$, which then decay through kinetic mixing with the photon to combinations of Standard Model leptons and mesons. The gamma-ray spectrum that results from such annihilations depends on the mass of the $\phi$. For $m_{\phi} \lsim 700$ MeV, final state radiation from charged leptons dominates, whereas decays of mesons dominate the gamma-ray spectrum in the case of $m_{\phi}\sim 1-2$ GeV (resulting in a spectrum similar to that found from dark matter candidates which annihilate to pairs of tau leptons). In either case, this gamma-ray spectrum provides a good fit to that observed from the Galactic Center. Furthermore, the $\phi$ decays to $e^+ e^-$ lead to synchrotron signals consistent with that observed from the Milky Way's radio filaments and diffusely throughout the Inner Galaxy (the ``WMAP Haze''). The normalization of each of these signals can be accommodated by a dark gauge coupling of $g_X \approx 0.06$, which also leads to a thermal relic abundance consistent with the measured cosmological abundance of dark matter.

Dark matter in this class of models is predicted to scatter elastically with protons, with a cross section that is determined by the mass of the dark force carrier and the degree of kinetic mixing between the force carrier and the photon, $\epsilon$. For $m_{\phi}\sim$ 100 MeV-3 GeV, values of $\epsilon\sim 10^{-3}-10^{-6}$ can lead to an elastic scattering rate capable of accounting for the signals reported by the DAMA/LIBRA, CoGeNT and CRESST-II collaborations. This range for $\epsilon$ is consistent with that naively expected from loop-suppressed processes. Lower values for $m_{\phi}$ and $\epsilon$ are not viable due to constraints from a combination of labortory experiments, supernovae, and the ellipticity of dark matter halos.

\bigskip

We would like to thank Matt Buckley, Pavel Fileviez Perez, Paddy Fox, Charles Gale, Patrick Meade, Guy Moore, Tracy Slatyer, and Kathryn Zurek for valuable discussions. DH is supported by the US Department of Energy. NW is supported by NSF grant \#0947827. WX is supported in part by a Schulich Fellowship.

\end{document}